\documentclass[aps,prb,twocolumn,showpacs,superscriptaddress]{revtex4}
\usepackage{amsmath}
\usepackage{latexsym}
\usepackage{amssymb}
\usepackage{graphicx}
\usepackage{bm}

\newcommand{\pdag}{{\phantom{\dagger}}}
\newcommand{\bq}{\begin{equation}}
\newcommand{\eq}{\end{equation}}
\newcommand{\bn}{\begin{eqnarray}}
\newcommand{\en}{\end{eqnarray}}

\begin{document}

\title{Inelastic cotunneling current and shot noise of an interacting
quantum dot with ferromagnetic correlations}

\author{Bing Dong}
\affiliation{Department of Physics, Shanghai Jiaotong University,
1954 Huashan Road, Shanghai 200030, China} 
\affiliation{Department of Physics and Engineering Physics, Stevens Institute 
of Technology,
Hoboken, New Jersey 07030, USA}

\author{X.L. Lei}
\affiliation{Department of Physics, Shanghai Jiaotong University,
1954 Huashan Road, Shanghai 200030, China}

\author{Norman J. M. Horing}
\affiliation{Department of Physics and Engineering Physics, Stevens
Institute of Technology, Hoboken, New Jersey 07030, USA}

\begin{abstract}

We explore inelastic cotunneling through a strongly Coulomb-blockaded
quantum dot attached to two ferromagnetic leads in the weak coupling limit 
using a generic quantum Langevin equation approach. We first develop a 
Bloch-type equation microscopically to describe the
cotunneling-induced spin relaxation dynamics, and then develop explicit 
analytical expressions for the local magnetization, current, and its 
fluctuations. On this basis, we predict a
novel zero-bias anomaly of the differential conductance in the absence of a 
magnetic field for the anti-parallel configuration, and asymmetric peak 
splitting in a magnetic field. 
Also, for the same system with large polarization, we find a negative 
zero-frequency differential shot noise in the low positive bias-voltage 
region. All these effects are ascribed to rapid spin-reversal due to 
underlying spin-flip cotunneling.

\end{abstract}


\today

\maketitle

\section{Introduction}

Recently, there has been extensive investigation of
spin-dependent tunneling in nanoscale semiconductor devices coupled to 
ferromagnetic leads\cite{Prinz}. It is well known that transport through a 
quantum dot (QD) includes two distinct mechanisms: sequential (first-order) 
tunneling if the QD level is in resonance with the Fermi levels of the 
electrodes, and higher-order tunneling (cotunneling) when the QD level is far 
removed from resonance, with sequential tunneling exponentially suppressed. 
Second-order cotunneling in the weak tunneling regime has already been 
experimentally
observed at temperatures above the Kondo temperature\cite{cotunnelingExp}. So 
far, most studies concerning spin-polarized tunneling through a QD have been 
focused on the sequential regime\cite{sequential1,sequential2} and on enhanced 
cotunneling in the strong tunneling regime, i.e. the
Kondo-type transport\cite{Martinek1,Pasupathy}. In particular, the theoretical
prediction of Kondo peak-splitting in a quantum dot in the presence of 
parallel (P)
spin-polarized leads\cite{Martinek1} has been confirmed in recent 
experiments\cite{Pasupathy}.

In this paper, we systematically analyze cotunneling through a strongly
Coulomb-blockaded QD attached to two ferromagnetic leads in the weak
tunneling regime at temperatures above the Kondo temperature. We develop an 
explicit expression for QD magnetization, which is believed to be of essential 
importance in
understanding the nonequilibrium Kondo effect, albeit that the formula is 
derived in second-order perturbation theory\cite{Parcollet,Paaske}. Our 
calculation of cotunneling current
predicts a novel zero-bias peak (ZBP) of differential conductance in the 
absence of a magnetic field, and an asymmetric peak splitting in a magnetic 
field for the anti-parallel (AP) configuration. Our discussion shows that 
these interesting features arise from spin-flip processes which induce rapid 
spin-reversal. An analogous ZBP in cotunneling was found in an earlier work by 
Weymann\cite{Weymann}. Here our explicit analytic expressions show similar 
results of the peak height and its temperature dependence as in 
Ref.~\onlinecite{Weymann}.

Moreover, we present an analytical investigation of zero-frequency shot noise 
in the cotunneling regime, which has been previously studied only in the 
sequential regime\cite{sequential1}. We predict that the differential shot 
noise at low bias-voltage region is heavily dependent on the polarization of 
electrodes in the AP configuration.

The remaining parts of the paper are arranged as follows. In Sec. II, we 
present the physical model and theoretical formulation used in this paper. We 
give the Bloch-type dynamical equations derived from a generic quantum 
Langevin equation approach\cite{Ackerhalt,Smirnov,Dong3} to describe the time 
evolution of the QD spin and all relevant correlation functions. In Sec. III, 
we discuss the nonequilibrium magnetization of the QD in detail and point out 
that the spin-flip cotunneling will induce a rapid spin accumulation in the AP 
configuration. It is our main result for theoretical explanation of the novel 
behavior of the differential conductance and zero-frequency shot noise, whose 
explicit expressions are derived based on the linear response theory in Sec. 
IV. Numerical evaluations and elaborate discussions are also provided in this 
section. Finally, a brief conclusions are given in Sec. V.        

\section{Theoretical Model and formulation}

Transport through an interacting QD with a
single energy level $\epsilon_{d}$ connected to two ferromagnetic leads is 
modeled by a single impurity Anderson Hamiltonian. To describe 
\emph{cotunneling} through the QD in the strongly Coulomb-blockaded regime 
(the on-site Coulomb interaction $U\rightarrow \infty$), in which the negative 
$\epsilon_d$ with $\epsilon_{d}\ll |\mu _{\eta}|$ ($\mu_{\eta }$ is the 
chemical potential in lead $\eta$ and the symmetrically applied bias voltage 
$V$ has $\mu_{L}=-\mu_{R}=eV/2$) is satisfied, so that charge fluctuations are 
completely suppressed, we perform a mapping of the Anderson Hamiltonian onto 
the subspace with one electron in the dot by a Schrieffer-Wolff unitary 
transformation\cite{Schrieffer}, i.e. onto the effective Kondo 
Hamiltonian\cite{Matveev,Dong3}:
\begin{align}
H=& \,H_{0}+H_{\mathrm{I}}, & &  \label{hamiltonian2} \\
H_{0}=& \,\sum_{\eta \mathbf{k}\sigma }\varepsilon _{\eta \mathbf{k}}c_{\eta
\mathbf{k}\sigma }^{\dag }c_{\eta \mathbf{k}\sigma }^{{\phantom{\dagger}}
}-\Delta S^{z},\cr 
H_{\mathrm{I}}= & \,\sum_{\eta ,\eta ^{\prime },\mathbf{k}
,\mathbf{k}^{\prime }}J_{\eta \eta ^{\prime }}\bigl [\bigl(c_{\eta \mathbf{k}
\uparrow }^{\dag }c_{\eta ^{\prime }\mathbf{k}^{\prime } 
\uparrow}^{{\phantom{\dagger}}} - c_{\eta \mathbf{k}\downarrow }^{\dag 
}c_{\eta ^{\prime }
\mathbf{k}^{\prime }\downarrow }^{{\phantom{\dagger}}}\bigr)S^{z}\cr&
\,+c_{\eta \mathbf{k}\uparrow }^{\dag }c_{\eta ^{\prime }\mathbf{k}^{\prime} 
\downarrow}^{{\phantom{\dagger}}}S^{-} + c_{\eta \mathbf{k}\downarrow}^{\dag} 
c_{\eta^{\prime} \mathbf{k}^{\prime }\uparrow }^{{\phantom{\dagger}}} 
S^{+}\bigr] + H_{\rm dir}, \cr
H_{\rm dir}= &\, J_{0} \sum_{\sigma} \bigl ( c_{L {\bf k} \sigma}^\dagger + 
c_{R {\bf k} \sigma}^\dagger \bigr ) \bigl ( c_{L {\bf k} \sigma}^\pdag + c_{R 
{\bf k} \sigma}^\pdag \bigr ),  \notag
\end{align}
where $c_{\eta \mathbf{k}\sigma }^{\dagger }$ ($c_{\eta \mathbf{k}\sigma }$)
is the creation (annihilation) operator for electrons with momentum 
$\mathbf{k}$, spin-$\sigma$ and energy $\epsilon _{\eta \mathbf{k}}$ in lead 
$\eta$ ($=\mathrm{L,R}$), $S^{z(\pm )}$ are Pauli spin operators of electrons 
in the QD, and $J_{\eta \eta ^{\prime }}$ is the Kondo exchange coupling 
constant. $H_{\rm dir}$ is the potential scattering term, which is decoupled 
from the electron spin due to the number of electrons in the dot level being 
one. As a result, this term has no influence on the dynamical evolution of the 
electron spin and behaves only as a direct bridge to connect the left and 
right leads and consequently to contribute a current being independent of the 
dynamics of the QD.
For an Anderson model with symmetrical coupling to the leads $t$, we have 
$J_{\eta \eta'}=2J_{0}=t^2/\epsilon_d$.      

In this paper, we consider the ferromagnetism of the leads by means of a 
spin-dependent, flat density of states, $\rho _{\eta \sigma }$: $\rho 
_{L\uparrow }=\rho _{R\uparrow}=(1+p)\rho _{0}$; $\rho _{L\downarrow }=\rho 
_{R\downarrow} = (1-p)\rho _{0}$
for the P configuration, and $\rho _{L\uparrow }=\rho _{R\downarrow}=(1+p)\rho 
_{0}$; $\rho_{L\downarrow }=\rho _{R\uparrow }=(1-p)\rho _{0}$
for the AP configuration, with the degree of spin polarization $p$ ($|p|\leq 
1$) for both leads.

The spin-splitting, $\Delta$, of the QD involves two contributions: an
ambient magnetic-field $B$-induced Zeeman term, $\Delta _{Z}=g\mu _{B}B$ and 
an effective
splitting, $\Delta _{\mathrm{P(AP)}}=D \sum_{\eta }J_{\eta \eta }(\rho _{\eta 
\uparrow} - \rho_{\eta \downarrow})=2pD(J_{LL}\pm J_{RR})\rho _{0},$ caused by 
ferromagnetic correlations of the two leads ($D$ is the band width of the 
leads). Based on this Hamiltonian, we find that (1) the differential 
conductance, $dI/dV$ of the cotunneling current exhibits double peaks with 
width $2\Delta_{p}$ as a function of voltage, $V,$ for P alignment of the lead 
magnetizations even without an ambient magnetic field; (2) the peak-splitting 
in the $dI/dV$ vs. $V$ curve without a
magnetic field is also predicted for the AP configuration only with asymmetric 
couplings $J_{LL}\neq J_{RR}$,
but it has a largely reduced width, $2\Delta _{\mathrm{AP}},$ in comparison 
with the P case; (3) in both cases the splitting can be removed by properly 
tuning the direction and
strength of the ambient magnetic field, to induce vanishing of the total 
spin-splitting $\Delta$. These results provide a good qualitative explanation 
of the experimental findings in Ref.~\onlinecite{Pasupathy}, and we can 
further clarify this issue in that the small but nonzero splitting found in 
the asymmetric AP case also stems from the lead-magnetization-induced 
spin-splitting in the QD.

We employ a generic Langevin equation approach to analyze this system in the 
weak cotunneling limit\cite{Ackerhalt,Smirnov,Dong3}. In our derivation, 
operators of the QD spin and the reservoirs are first expressed formally by 
integration of their Heisenberg equations of motion (EOM), exactly to all 
orders of $J_{\eta\eta'}$. Next, under the assumption that the time scale of 
decay processes is much slower than that of free evolution, we replace the 
time-dependent
operators involved in the integrals of these EOM's approximately in terms of 
their free
evolutions. Thirdly, these EOM's are expanded in powers of $J_{\eta\eta'}$ up 
to second
order. To this end, a Bloch-type dynamical equation is established to describe 
the time evolution of the QD spin variable as,
\bq 
\dot S^{z}= -4C_{\rm P(AP)}(\Delta) \, S^{z} + 2R_{\rm
P(AP)}(\Delta), \label{sz} 
\eq 
in which
\begin{subequations}
\bn
C_{\rm P}(\omega)&=& \frac{\pi}{2} \left ( g_{LL}+ g_{RR} \right ) (1-p^2) T 
\varphi \left ( \frac{\omega}{T} \right ) \cr
&& \hspace{-1.5cm}+ \frac{\pi}{2} g_{LR} (1-p^2) T \left [ \varphi \left ( 
\frac{\omega + V}{T}\right ) + \varphi \left ( \frac{\omega- V}{T}\right ) 
\right
], \label{Cp} \\
R_{\rm P}(\omega)&=& \frac{\pi}{2} \left ( g_{LL}+ g_{RR} + 2g_{LR} \right ) 
(1-p^2) \omega, \en
\end{subequations}
for the P configuration, and
\begin{subequations}
\bn
C_{\rm AP}(\omega)&=& \frac{\pi}{2} \left ( g_{LL}+ g_{RR} \right ) (1-p^2) T 
\varphi \left ( \frac{\omega}{T} \right ) + \frac{\pi}{2} g_{LR} T\cr
&& \hspace{-1.5cm} \times \left [ (1+p)^2 \varphi \left ( \frac{\omega + 
V}{T}\right )  + (1-p)^2 \varphi \left ( \frac{\omega- V}{T}\right ) \right ], 
\label{Cap} \\
R_{\rm AP}(\omega)&=& \frac{\pi}{2} \left ( g_{LL}+ g_{RR} \right )
(1-p^2) \omega \cr && + \pi g_{LR} (1+p^2) \omega + 2\pi g_{LR} p V,
\en
\end{subequations}
for the AP configuration, with $g_{\eta \eta ^{\prime }}\equiv J_{\eta 
\eta^{\prime}}^{2} \rho_{0}^{2}$ and $\varphi (x)\equiv x\coth (x/2)$. Once 
again, we observe that the direct tunneling term has no contribution to 
dissipation of the QD spin variables. Throughout, we will use units with 
$\hbar =k_{B}=e=1$. 

Based on the dynamical equation, Eq.~(\ref{sz}), we identify the 
cotunneling-induced
spin relaxation rate as $1/T_{1}^{\mathrm{P(AP)}}=4C_{\mathrm{P(AP)}}(\Delta 
)$, which stems
completely from spin-flip events involving both single-barrier processes 
[Fig.~3(c)-(h) in Ref.~\onlinecite{Dong3}] and double-barrier processes [i.e. 
electron-transferring (ET) processes] [Fig.~3(m)-(p) in 
Ref.~\onlinecite{Dong3}]. For the case of fully spin-polarized leads ($p=\pm 
1$) in the P alignment, cotunneling is unable to flip the electron spin in the 
QD, thus implying that there is no spin relaxation, i.e. 
$1/T_{1}^{\mathrm{P}}=0$ [$C(R)_{\mathrm{P}}=0$ at $p=\pm 1$
from Eq.~(\ref{Cp})]. For this case, we also have $\dot{S}^{z}=0$, meaning 
that the QD spin orientation remains unchanged from its initial state. In 
contrast to this, in the case of fully AP spin-polarized leads, spin 
relaxation rate is nonzero and it arises solely from ET events [Fig.~1(c) and 
(d) below].

\section{Nonequilibrium Magnetization}

The nonequilibrium local magnetization, $S_{\mathrm{P(AP)}}^{z\infty }$, of 
the QD in P(AP) configuration is readily obtained using the steady-state 
solution of Eq.~(\ref{sz}) as
\begin{widetext}
\bn
S_{\rm P}^{z\infty}(\Delta,V)&=&\frac{R_{\rm P}(\Delta)}{2C_{\rm P}(\Delta)} 
=\frac{\displaystyle \left ( \frac{g_{LL} + g_{RR}}{2} + g_{LR} \right ) 
\frac{\Delta}{T}}{\displaystyle \left ( g_{LL} + g_{RR} \right ) \varphi \left 
( \frac{\Delta}{T} \right ) + g_{LR} \left [ \varphi \left ( \frac{\Delta + 
V}{T}\right ) + \varphi \left ( \frac{\Delta- V}{T}\right ) \right ]}, 
\label{magnetizationP} \\
S_{\rm AP}^{z\infty}(\Delta,V)&=&\frac{R_{\rm AP}(\Delta)}{2C_{\rm 
AP}(\Delta)} =\frac{\displaystyle \frac{g_{LL} + g_{RR}}{2} (1-p^2) 
\frac{\Delta}{T}+ g_{LR} (1+p^2) \frac{\Delta}{T} + 2g_{LR} p 
\frac{V}{T}}{\displaystyle \left ( g_{LL} + g_{RR} \right ) (1-p^2) \varphi 
\left ( \frac{\Delta}{T} \right ) + g_{LR} \left [ (1+p)^2 \varphi \left ( 
\frac{\Delta + V}{T}\right ) + (1-p)^2 \varphi \left ( \frac{\Delta- 
V}{T}\right ) \right ]}. \label{magnetizationAP}
\en
\end{widetext}
The P magnetization formula, Eq.~(\ref{magnetizationP}), is
identical with previous theoretical results for normal 
leads\cite{Parcollet,Paaske,Dong3}, and it is worth noting that
the polarization of the reservoirs is fully reflected in the effective
spin-splitting, $\Delta _{\mathrm{P}}$. If an ambient magnetic field is
applied to induce vanishing of the spin-splitting, $\Delta =0$, we have 
$S_{\mathrm{P}}^{z\infty }(0,V)=0$.

However, the local magnetization in the AP configuration is significantly
different. Our calculated results are shown as functions of bias-voltage in 
Fig.~1. The relevant parameters in our calculations are: 
$g_{LL}=g_{RR}=g_{LR}=g=0.25 \times 10^{-2}$; $D=10$ (corresponding to Kondo 
temperature $T_{K}=4.5\times 10^{-5}D$ and $\Delta _{\mathrm{P}}=0.2pD$), and 
the temperature is $T/D=0.002$. In the absence of an external magnetic field, 
$S_{\mathrm{AP}}^{z\infty }$ is a two-valued
function, $\pm p/(1+p^{2})$ if $|V|/T\gg 1,$ and zero net spin of the QD 
occurs only at $V=0$ when $p\neq 0$ [Fig.~1(a)]. Nonzero local spin
is generated by the application of a bias-voltage, indicating spin 
accumulation in the QD under nonequilibrium conditions, and its orientation is 
rapidly flipped when the direction of the bias-voltage is reversed. This 
interesting behavior can be physically explained by \emph{spin-flip elastic 
cotunneling} processes, as shown in Figs.~1(c) and (d). If
$V>0$ [Fig.~1(c)], the dominant cotunneling process is: first a spin-down 
electron in the QD tunnels out to
the right lead, then a spin-up majority electron successively enters into the 
QD. This process continually pumps spin-down electrons out of the QD, while 
injecting spin-up electrons into the QD, giving rise to an up-spin pileup, 
until a steady-state is reached. On the contrary, for $V<0$, the reverse 
process takes place, causing a down-spin accumulation [Fig.~1(d)]. Another
interesting consequence of this spin-flip cotunneling is the appearance of a
sharp peak at $V=0$ in the $\frac{\partial S^{z}}{\partial V}$ vs. $V$ curves, 
which leads
to a novel ZBP in differential conductance for the AP configuration, as shown 
in Fig.~2(b). Simple calculation shows that the peak height of $\frac{\partial 
S^{z}}{\partial V}$ at $V=0$ is proportional to the polarization $p$ of 
electrodes, but is inversely proportional to the temperature $T$:
\bq
\frac{\partial S_{\rm AP}^{z}}{\partial V}{\Big |}_{V\rightarrow 
0}=\frac{p}{4T}. \label{sappv}
\eq
While for $|V|/T\rightarrow \infty$, we have $\frac{\partial S_{\rm 
AP}^{z}}{\partial V}=0$. These two relations are useful in analyzing ZBP 
height of differential conductance. The temperature behavior of 
$\frac{\partial S^{z}}{\partial V}$ is shown schematically in Fig.~3(a) in the 
following section.   

\begin{figure}[htb]
\begin{center}
\includegraphics [width=8.5cm,height=3.5cm,angle=0,clip=on]{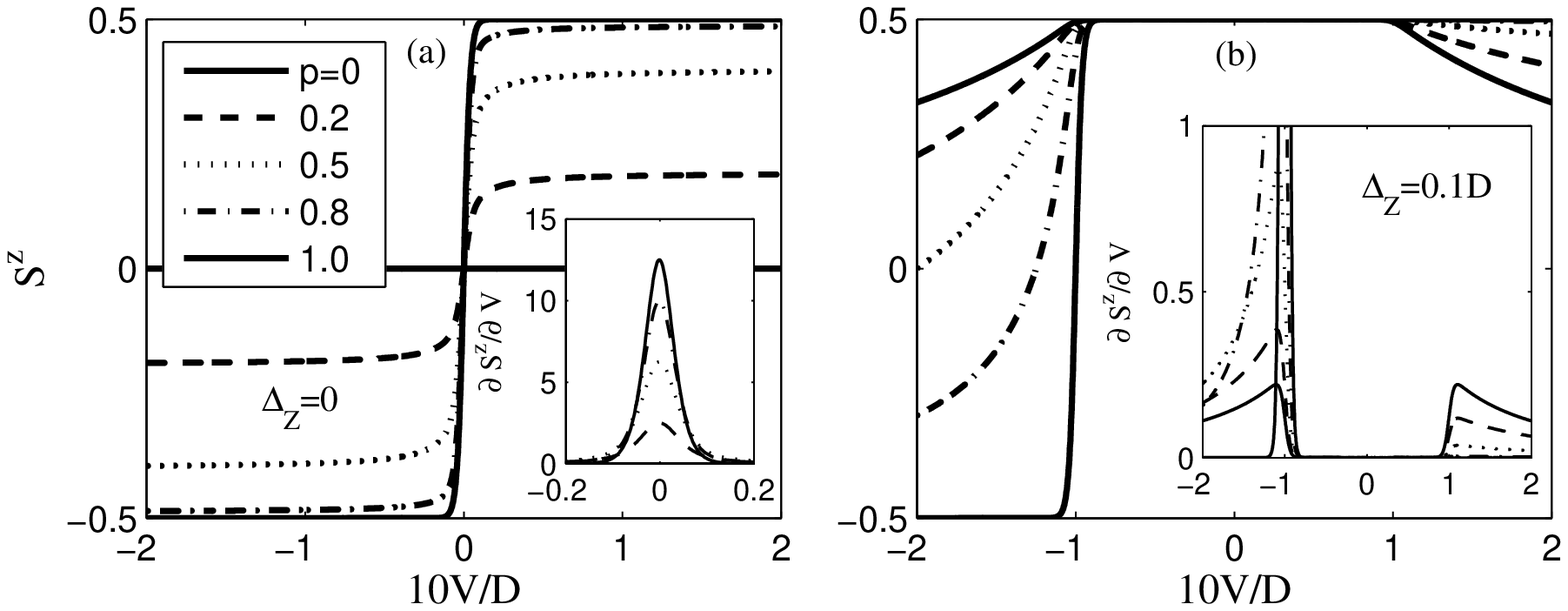} 
\includegraphics[width=7.cm,height=3.cm,angle=0,clip=on]{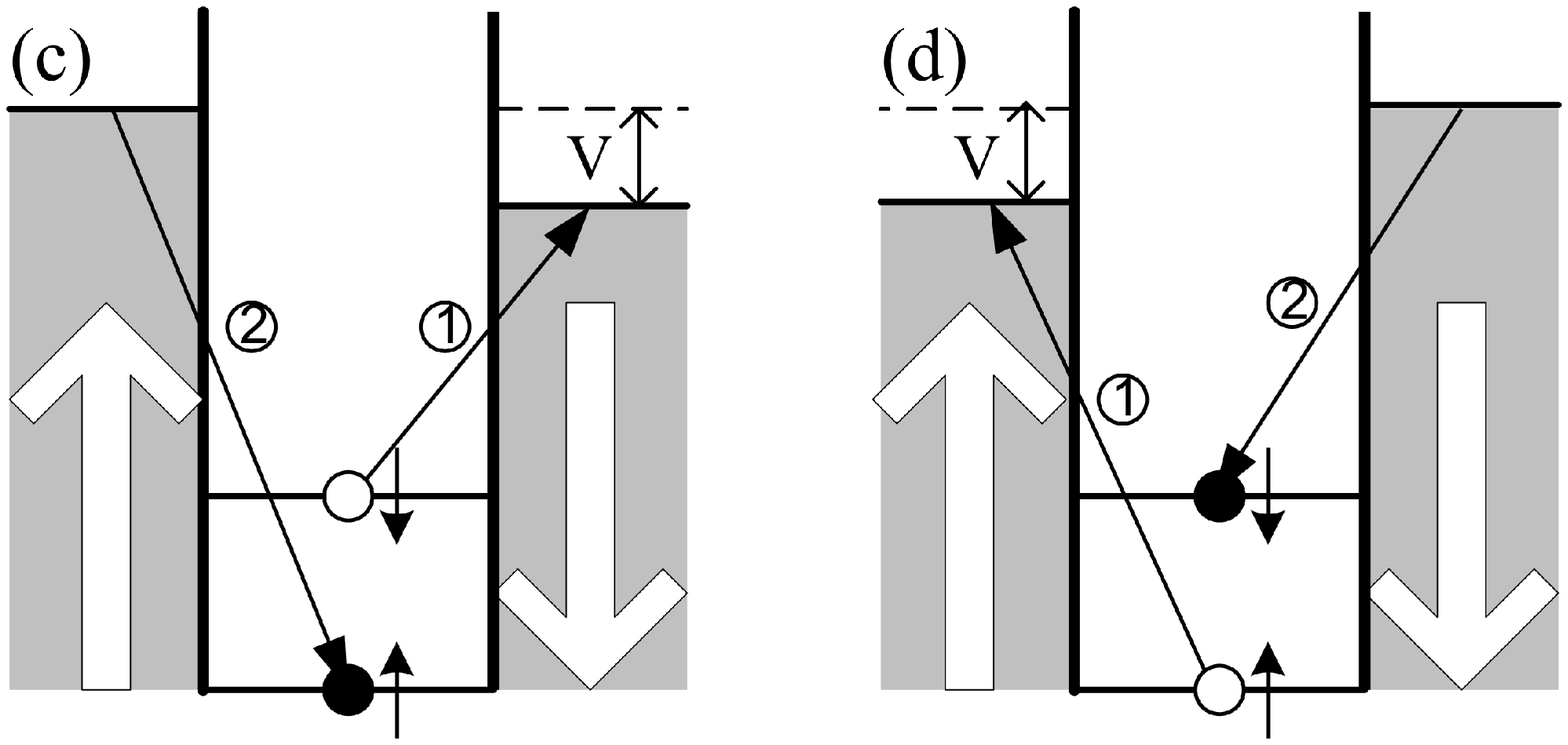}
\end{center}
\caption{Nonequilibrium QD magnetization, $S^z$, vs. bias-voltage, $V$, for
the AP configuration at (a) $\Delta_Z=0$, and (b) $\Delta_Z\neq 0$. Insets: 
$\frac{\partial S^z}{\partial V}$ vs. $V$ curves. (c) and (d) ET spin-flip
cotunneling processes.}
\end{figure}

A finite ambient magnetic field splits the peak in $\frac{\partial 
S^{z}}{\partial V}$ into two peaks at $V=\pm \Delta $ with unequal heights 
[Fig.~1(b)]. This
occurs because spin-flip cotunneling becomes an energy-consuming 
(\emph{inelastic}) process in a nonzero magnetic field, thus requiring a 
sufficiently strong external bias-voltage for
activation. According to Eq.~(\ref{magnetizationAP}), the spin-reversal takes 
place approximately at $\Delta /V=-p$. Moreover, the application of a positive 
magnetic field makes the cotunneling shown in Fig.~1(c) prevail over that in 
Fig.~1(d). Correspondingly, the peak at $V=\Delta $ is suppressed, but the 
peak at $V=-\Delta $ is significantly enhanced if $p>0$. The spin reversal 
property discussed above has potential for application in quantum computing 
and quantum information processing. Finally, in the absence of a bias voltage,
the local magnetization reduces to the equilibrium expression 
$S_{\mathrm{P(AP)}}^{z\infty}(\Delta ,0)=\frac{1}{2}\tanh(\Delta /2T)$.

\section{Current and Shot Noise}

The tunneling current operator of the QD
is defined as the time rate of change of charge density $N_{\eta 
}=\sum_{\mathbf{k}
\sigma }c_{\eta \mathbf{k}\sigma }^{\dagger }c_{\eta 
\mathbf{k}\sigma}^{{\phantom{\dagger}}}$ in lead $\eta $: $J_{\eta 
}(t)=\dot{N}_{\eta }$. From linear-response theory we have
\begin{equation}
I=\langle J_{\eta }(t)\rangle =-i\int_{-\infty }^{t}dt^{\prime }\langle
\lbrack J_{\eta }(t),H_{\mathrm{I}}(t^{\prime })]_{-}\rangle _{0},
\label{current}
\end{equation}
where the statistical average $\langle \cdots \rangle _{0}$ is performed
with respect to two decoupled subsystems, QD and reservoirs.

Nonequilibrium quantum shot noise is another active subject in mesoscopic
physics, because the current correlation function can provide further 
information about electronic correlations which is not available from 
conductance probes alone\cite{Blanter,Dong2}. The noise spectrum is defined as 
the Fourier transform of the current-current correlation function, $S_{\eta 
\eta^{\prime }}(\tau )$, and it too can be calculated using linear-response 
theory:
\begin{equation}
S_{\eta \eta ^{\prime }}(\omega )=\int_{-\infty }^{\infty }d\tau e^{i\omega
\tau }\frac{1}{2}\langle \lbrack \delta J_{\eta }(t),\delta J_{\eta ^{\prime
}}(t^{\prime })]_{+}\rangle _{0},  \label{def:sn}
\end{equation}
with $\delta J_{\eta }(t)=J_{\eta }(t)-\langle J_{\eta }(t)\rangle $.

Considering that the interaction Hamiltonian $H_{\rm I}$ in 
Eq.~(\ref{hamiltonian2}) contains two components, the direct tunneling term 
$H_{\rm dir}$ (the direct tunneling channel) and the cotunneling term (the 
indirect tunneling channel), we can divide the current and shot noise into 
three contributions,
\begin{subequations}
\bn
I&=& I^{0} + I^{\rm d} + I^{\rm in},\\
S_{LL}(\omega)&=& S_{LL}^{\rm 0}(\omega) + S_{LL}^{\rm d}(\omega) + 
S_{LL}^{\rm in}(\omega),
\en
\end{subequations}
due to the indirect channel, direct channel, and the interference effect 
between the two channels, respectively.   
After lengthy but straightforward calculations, we obtained explicit
expressions for steady-state cotunneling current and for frequency-independent 
auto-correlation shot noise via the indirect channel:
\begin{subequations}
\label{Pis} 
\bn 
I_{\rm P}^{0}/\pi g_{LR}&=& (1+p^2) V + 2(1-p^2) V + 2 T
S_{\rm P}^{z\infty} \cr
&& \hspace{-1.5cm} \times (1-p^2) \bigl [ \varphi \bigl ( \frac{V-\Delta}{T} 
\bigr ) - \varphi \bigl ( \frac{V+\Delta}{T} \bigr )\bigr ], \label{ip} \\
S_{LL}^{\rm P,0}(0)/\pi g_{LR}&=& (1+p^2) T \varphi \bigl (
\frac{V}{T}\bigr ) + (1-p^2) T \cr
&& \hspace{-2.5cm} \times \bigl [ \varphi \bigl ( \frac{V-\Delta}{T} \bigr ) + 
\varphi \bigl ( \frac{V+\Delta}{T} \bigr ) \bigr ] - 4 S_{\rm P}^{z\infty} 
(1-p^2)
\Delta , \label{sp} 
\en
\end{subequations}
for the P configuration, and
\begin{subequations}
\label{APis} 
\bn 
I_{\rm AP}^{0}/\pi g_{LR}&=& (1-p^2) V + [(1-p)^2
(V-\Delta) \cr && \hspace{-2.5cm} + (1+p)^2(V+\Delta)] + 2 T S_{\rm
AP}^{z\infty} \cr
&& \hspace{-2.5cm} \times \bigl [ (1-p)^2 \varphi \bigl ( \frac{V-\Delta}{T} 
\bigr ) - (1+p)^2 \varphi \bigl ( \frac{V+\Delta}{T} \bigr )\bigr ], 
\label{iap} \\
S_{LL}^{\rm AP,0}(0)/\pi g_{LR}&=& (1-p^2) T \varphi \bigl (
\frac{V}{T}\bigr ) \cr
&& \hspace{-2.5cm} + T \bigl [ (1-p)^2 \varphi \bigl ( \frac{V-\Delta}{T} 
\bigr ) + (1+p)^2 \varphi \bigl ( \frac{V+\Delta}{T} \bigr ) \bigr ] \cr
&& \hspace{-2.5cm} + 2 S_{\rm AP}^{z\infty} [(1-p)^2 (V-\Delta) - (1+p)^2 
(V+\Delta)], \label{sap} 
\en
\end{subequations}
for the AP configuration. Further elaborate analyses clarify that the first
linearly bias-voltage-dependent terms in Eqs.~(\ref{Pis}) and (\ref{APis}) 
result from spin-conservative ET cotunneling processes, whereas spin-flip ET 
events are responsible for the other contributions\cite{Dong3}. It is 
noteworthy that the first terms on
the right hand sides of Eq.~(\ref{ip}) [(\ref{iap})] and Eq.~(\ref{iap}) 
[(\ref{sap})] obey the nonequilibrium fluctuation-dissipation (NFD) relation, 
while the other two terms represent the generalized NFD relation due to 
energy-consumption cotunneling processes\cite{NFDT}. The direct tunneling 
yields\cite{directtunneling}:
\begin{subequations}
\bn
I_{\rm P}^{\rm d}/\pi g_{LR} &=& 4\frac{g_0}{g_{LR}} (1+p^2) V, \\   
S_{LL}^{\rm P, d}(0)/\pi g_{LR} &=& 4\frac{g_0}{g_{LR}} (1+p^2) T\varphi \bigl 
( \frac{V}{T}\bigr ),
\en   
\end{subequations}
for the P configuration, and
\begin{subequations}
\bn
I_{\rm AP}^{\rm d}/\pi g_{LR} &=& 4\frac{g_0}{g_{LR}} (1-p^2) V, \\   
S_{LL}^{\rm AP, d}(0)/\pi g_{LR} &=& 4\frac{g_0}{g_{LR}} (1-p^2) T\varphi 
\bigl ( \frac{V}{T} \bigr ),
\en   
\end{subequations}
for the AP configuration with $g_0=J_0^2 \rho_0^2=g_{LR}/4$. Because the 
direct tunneling is a spin-conservative process, it only interfere with the 
non-spin-flip cotunneling processes. The interference effect thus contributes 
to the current and shot noise only for the P configuration:
\begin{subequations}
\bn
I_{\rm P}^{\rm in}/\pi g_{LR} &=& 16 \sqrt{\frac{g_0}{g_{LR}}} p S_{\rm 
P}^{z\infty} V, \\   
S_{LL}^{\rm P, in}(0)/\pi g_{LR} &=& 16 \sqrt{\frac{g_0}{g_{LR}}} p S_{\rm 
P}^{z\infty} T\varphi \bigl ( \frac{V}{T}\bigr ).
\en   
\end{subequations}
Setting $p=0$, we obtain the cotunneling current formula for the 
normal-lead/QD/normal-lead system [Eq.~(36) in 
Ref.~\onlinecite{Dong3}]\cite{DongE}, and the corresponding 
frequency-independent shot noise $S_{LL}(0)$ is given by 
[$S^{z\infty}=S^{z\infty}_{\rm P}$]:
\bn
S_{LL}(0)/\pi g_{LR} &=& 2T \varphi \bigl (
\frac{V}{T}\bigr ) -4 S^{z\infty} \Delta \cr
&& + T \bigl [ \varphi \bigl ( \frac{V-\Delta}{T} \bigr ) +  \varphi \bigl ( 
\frac{V+\Delta}{T} \bigr ) \bigr ].
\en

\begin{figure}[htb]
\includegraphics [width=8.5cm,height=8.cm,angle=0,clip=on]{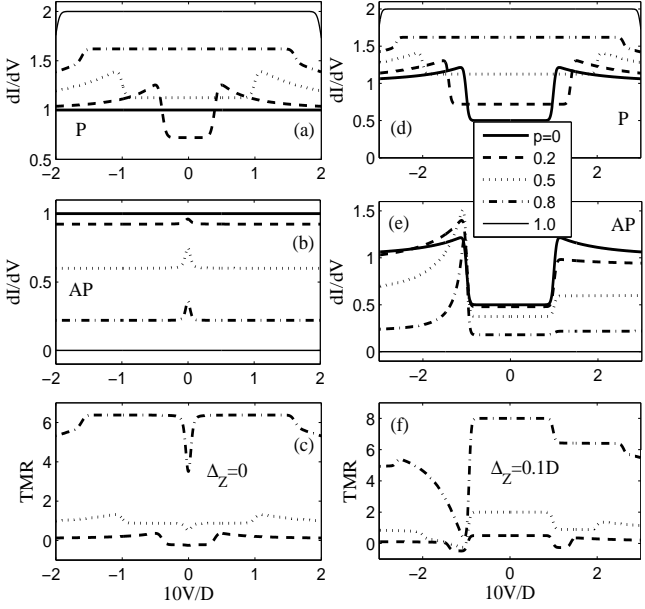} 
\caption{Differential conductances and TMRs as functions of bias-voltage
under zero (a)-(c) and nonzero (d)-(f) external magnetic fields for various
polarizations $p$. The other parameters are the same as in Fig.~1.}
\label{fig2}
\end{figure}

We plot the differential conductance, 
$G_{\mathrm{P(AP)}}=\frac{dI_{\mathrm{P(AP)}}}{dV}$ (in units of $4\pi g$), 
and the tunnel magnetoresistance, 
TMR=$(G_{\mathrm{P}}-G_{\mathrm{AP}})/G_{\mathrm{AP}}$, vs. bias-voltage in 
Fig.~2. For the P alignment, we find a characteristic jump
at $V=\pm \Delta $ even at zero ambient magnetic field, $\Delta _{Z}=0$, due 
to the lead
polarization-induced spin-splitting, $\Delta _{\mathrm{P}}$. But for 
spin-polarized leads with larger polarization, $p=0.8$ and $p=1$ in Fig.~2(a), 
the peak-splitting disappears, as can be understood from the fact that the
spin-flip process can hardly ($p=0.8$) or fully not ($p=1$) occur in these 
specific cases, thus no additional channel is opened for ET even when $V\geq 
\Delta$, in comparison with smaller polarization $p<0.8$\cite{Dong3}. For the 
case of vanishing spin-splitting $\Delta=0$, the linear conductance, $G_{\rm 
P}^0=\frac{dI_{\rm P}}{dV}|_{V\rightarrow 0}$, is readily given by
\bq
G_{\rm P}^0/\pi g =4. \label{gp0}
\eq

On the contrary,
for the AP configuration, a ZBP emerges. Obviously, this behavior stems
mathematically from the sharp peak in $\frac{\partial S^{z}}{\partial V}$ 
around $V=0$
[Fig.~1(a)] according to Eq.~(\ref{iap}). Consequently, one can conclude that 
the underlying rapid spin-reversal due to \emph{spin-flip elastic} cotunneling 
is responsible for the novel ZBP from a physical point of view. With the help 
of Eq.~(\ref{sappv}), we obtain the linear conductance, $G_{\rm 
AP}^0=\frac{dI_{\rm AP}}{dV}|_{V\rightarrow 0}$, as
\bq
G_{\rm AP}^0/\pi g=3+p^2-16Tp \frac{\partial S_{\rm AP}^z}{\partial V}{\Big 
|}_{V\rightarrow 0}=4(1-p^2). \label{gap0}
\eq
We can also calculate the differential conductance $G_{\rm AP}$ at the limit 
of $V/T\rightarrow \infty$:
\bq
G_{\rm AP}^{\infty}/\pi g= 3+p^2-8p S_{\rm AP}^z|_{V\rightarrow\infty}= 
\frac{4(1-p^2)}{1+p^2}. \label{gapi}
\eq
It is easy to see that $G_{\rm AP}^0>G_{\rm AP}^\infty$ is always satisfied, 
which reflects the ZBP. Moreover, the relative height $x$ of the ZBP is 
defined as
\bq
x=\frac{G_{\rm AP}^0 - G_{\rm AP}^\infty}{G_{\rm AP}^{\infty}}=p^2,
\eq
which is exactly the same as the result in Ref.~\onlinecite{Weymann}. 
Furthermore, we examine the temperature dependence of the ZBP as shown in 
Fig.~3(b). It is found that increasing temperature will enhance the width of 
the ZBP, but reduce the peak height a bit, which is qualitatively consistent 
with the numerical calculations in Ref.~\onlinecite{Weymann} [see their 
Fig.~1(b)]. Interestingly, the novel ZBP in the AP configuration is robust 
over a rather wide region of temperature. It is worth emphasizing that our 
investigation is valid for the cotunneling through a strongly 
Coulomb-blockaded QD ($U\rightarrow \infty$) in the weak-tunneling limit, thus 
the contribution of the charge fluctuation effect is completely excluded in 
the present analysis. We believe that this is the reason for the difference 
between our analysis and the numerical calculations of 
Ref.~\onlinecite{Weymann}.             

\begin{figure}[htb]
\includegraphics [width=8.5cm,height=3.5cm,angle=0,clip=on]{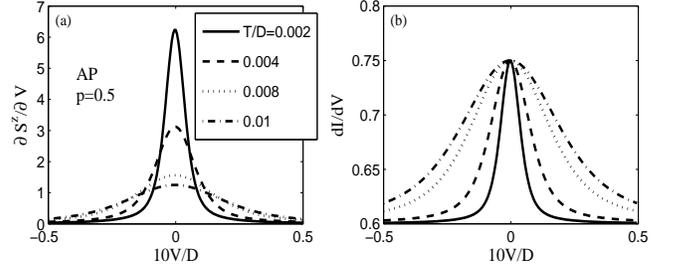} 
\caption{Temperature dependences of $\frac{\partial S^{z}}{\partial V}$ (a) 
and differential conductance (b) as functions of
bias-voltage in the AP configuration with zero magnetic field and $p=0.5$.} 
\label{fig3}
\end{figure}

In combination with the peak-splitting in the P configuration, the ZBP in the 
AP configuration will cause a deep dip in TMR at $V=0$, as shown in Fig.~2(c). 
Analogously, an ambient magnetic field leads to
peak-splitting since \emph{inelastic spin-flip} cotunneling requires 
sufficient energy for
activation. Contrary to the P alignment case, the two peaks in 
$\frac{dI_{\mathrm{AP}}}{dV}$ have unequal heights due in part to the 
asymmetry in $\frac{\partial S^{z}}{\partial V}$ with nonzero $\Delta_{Z}$ 
[Fig.~1(b)]. Correspondingly, the magnetic field changes the behavior of TMR 
[Fig.~2(f)].

\begin{figure}[htb]
\includegraphics [width=8.5cm,height=7.cm,angle=0,clip=on]{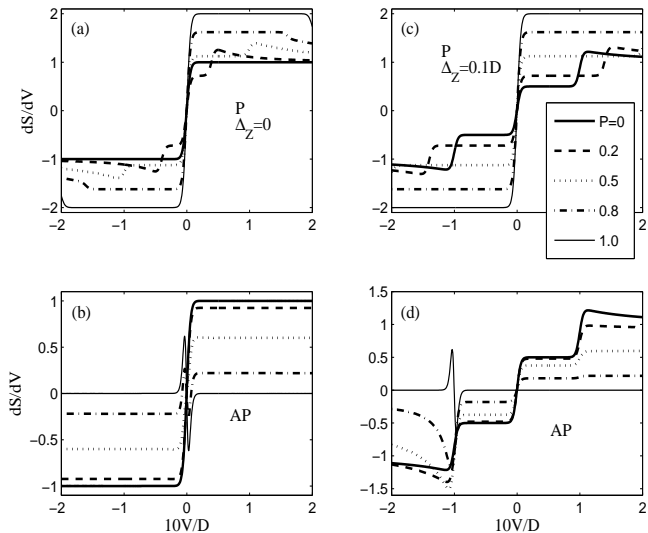} 
\caption{Differential zero-frequency shot noise as a function of
bias-voltage with zero (a), (b) and nonzero (c), (d) external
magnetic fields for various polarizations $p$. The other parameters are the 
same as in Fig.~1.} \label{fig4}
\end{figure}

Richly detailed structures are observed in the differential zero-frequency
shot noise, $\frac{dS}{dV}$, as shown in Fig.~4. Differing from the 
differential
conductance, the resulting QD spin-splitting causes an additional step 
structure in $\frac{dS}{dV}$ with equal (P) and unequal (AP) heights 
[Fig.~4(a), (c), and (d)]. Simple algebraic calculation gives:
\bn
&&\frac{dS^{\rm P}}{dV}{\Big |}_{V\rightarrow 0}=\frac{dS^{\rm AP}}{dV}{\Big 
|}_{V\rightarrow 0}=0,\\
&&\frac{dS^{\rm P}}{dV}{\Big |}_{V\rightarrow \infty}=4 > 0,\\
&&\frac{dS^{\rm AP}}{dV}{\Big |}_{V\rightarrow \infty} = 
\frac{4(1-p^2)}{1+p^2} \geq 0,
\en  
at $\Delta=0$. Generally, the zero-frequency shot noise will always increase 
with increasing bias-voltage. However, a contrary phenomenon is predicted in 
our calculation for the case of the AP alignment leads with large polarization 
$p$: the differential shot noise changes sign near $V=-\Delta$. This behavior 
can also be ascribed to the rather sharp peak in $\frac{\partial 
S^{z}}{\partial V}$ due to large $p$, i.e. to rapid QD spin-reversal. In 
Fig.~5, we also examine the temperature dependence of the differential shot 
noise at $\Delta=0$. It is seen that this behavior is more pronounced for 
higher temperature. 

\begin{figure}[htb]
\includegraphics [width=8.5cm,height=3.5cm,angle=0,clip=on]{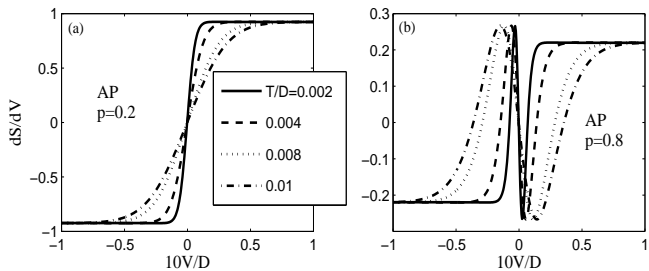} 
\caption{Temperature dependencies of differential shot noise as functions of 
bias-voltage in the AP configuration without external magnetic field for 
$p=0.2$ (a) and $p=0.8$ (b).} \label{fig5}
\end{figure}

\section{Conclusions}

In summary, we have applied a quantum Langevin
equation approach to derive a Bloch-type equation microscopically for 
analytical studies of inelastic cotunneling, which we have carried out in 
detail, determining the local magnetization, current and its fluctuations, in 
a single QD attached to two ferromagnetic
electrodes. Our studies reveal a number of interesting novel characteristics 
intimately related to \emph{spin-flip} processes in the AP configuration: 1) a 
ZBP of the differential conductance at zero external magnetic field; 2) 
asymmetric peak-splitting in the presence of a nonzero magnetic field; and 3) 
decreasing shot noise near $V=-\Delta$. In addition, we have found that these 
characteristics are robust against temperature.

\begin{acknowledgments}

This work was supported by Projects of the National Science Foundation of 
China, the Shanghai Municipal Commission of Science and Technology, the 
Shanghai Pujiang Program, and Program for New Century Excellent Talents in 
University (NCET). NJMH is supported by the DURINT program administered by the 
US Army Research Office, DAAD Grant No.19-01-1-0592.

\end{acknowledgments}


\begin{thebibliography}{99}

\bibitem{Prinz}{G.A. Prinz, Science \textbf{282}, 1660 (1998).}

\bibitem{cotunnelingExp}  {S.De Franceschi, S. Sasaki, J.M. Elzerman, W.G.
van der Wiel, S. Tarucha, and L.P. Kouwenhoven, Phys. Rev. Lett. \textbf{86}, 
878 (2001); A.
Kogan, S. Amasha, D. Goldhaber-Gordon, G. Granger, M.A. Kastner, and H. 
Shtrikman, Phys. Rev.
Lett. \textbf{93}, 166602 (2004); D.M. Zumb\"{u}hl, C.M. Marcus, M.P. Hanson 
and A.C. Gossard,
Phys. Rev. Lett. \textbf{93}, 256801 (2004).}

\bibitem{sequential1}  {B.R. Bulka, Phys. Rev. B \textbf{62}, 1186 (2000);
A. Cottet, W. Belzig, and C. Bruder, Phys. Rev. Lett. \textbf{92}, 206801 
(2004); I. Djuric, B.
Dong, H.L. Cui, IEEE Trans. on Nano. \textbf{4}, 71 (2005)}

\bibitem{sequential2}  {W. Rudzi\'{n}ski and J. Barna\'{s}, Phys. Rev. B
\textbf{64}, 85318 (2001); B. Dong, H.L. Cui, and X.L. Lei, Phys. Rev. B 
\textbf{69}, 35324 (2004).}

\bibitem{Martinek1}  {J. Martinek, Y. Utsumi, H. Imamura, J. Barna\'{s}, S.
Maekawa, J. K\"{o}nig, and G. Sch\"{o}n, Phys. Rev. Lett. \textbf{91}, 127203 
(2003); B. Dong, H.L.
Cui, S.Y. Liu, and X.L. Lei, J. Phys.: Condens. Matter \textbf{15}, 8435 
(2003); J. Martinek,
M. Sindel, L. Borda, J. Barna\'{s}, J. K\"{o}nig, G. Sch\"{o}n, and J. von 
Delft, Phys. Rev.
Lett. \textbf{91}, 247202 (2003); M.S. Choi, D. S\'{a}nchez, R. L\'{o}pez, 
Phys. Rev. Lett. \textbf{92}, 56601 (2004).}

\bibitem{Pasupathy}  {A.N. Pasupathy, R.C. Bialczak, J. Martinek, J.E.
Grose, L.A.K. Donev, P.L. McEuen, and D.C. Ralph, Science \textbf{306}, 86 
(2004); J. Nyg{\aa}rd, W.F. Kochl, N. Mason, L. DiCarlo, and C.M. Marcus, 
cond-mat/0410467.}

\bibitem{Parcollet}{O. Parcollet and C. Hooley, Phys. Rev. B \textbf{66},
85315 (2002).} 

\bibitem{Paaske}{J. Paaske, A. Rosch, and P. W\"{o}lfle, Phys. Rev. B 
\textbf{69}, 155330 (2004).}

\bibitem{Weymann}{I. Weymann, J. Barna\'s, J. K\"onig, J. Martinek, and G. 
Sch\"on, Phys. Rev. B {\bf 72}, 113301 (2005).}

\bibitem{Ackerhalt}{J.R. Ackerhalt and J.H. Eberly, Phys. Rev. D {\bf 10}, 
3350 (1974).}

\bibitem{Smirnov}{G.F. Efremov and A.Yu. Smirnov, Zh. \'Eksp. Teor. Fiz. {\bf 
80}, 1071 (1981) [Sov. Phys. JETP {\bf 53}, 547 (1981)].}

\bibitem{Dong3}{B. Dong, N.J.M. Horing, and H.L. Cui, Phys. Rev. B {\bf 72}, 
165326 (2005).}

\bibitem{Schrieffer}  {J.R. Schrieffer and P.A. Wolff, Phys. Rev. 
\textbf{149}, 491 (1966).}

\bibitem{Matveev}{K.A. Matveev, Zh. Eksp. Teor. Fiz. \textbf{99}, 1598
(1991) [Sov. Phys. JETP \textbf{72}, 892 (1991)]; A. Furusaki and K.A. 
Matveev, Phys. Rev. B \textbf{52}, 16676 (1995).}

\bibitem{Blanter}  {For a review, see Ya.M. Blanter and M. B\"{u}ttiker,
Phys. Rep. \textbf{336}, 1 (2000).}

\bibitem{Dong2}{B. Dong and X.L. Lei, J. Phys.: Condens. Matter \textbf{14}, 
4963 (2002).}

\bibitem{NFDT}{D. Rogovin and D.J. Scalapino, Ann. Phys. (N.Y.) \textbf{86}, 1 
(1974); E.V. Sukhorukov, G. Burkard, and D. Loss, Phys. Rev. B \textbf{63}, 
125315 (2001).}

\bibitem{directtunneling}{The calculations for the contributions of the direct 
tunneling and the interference effect to current are easy and straightforward. 
The interesting reader could refer to our recent paper, Bing Dong, N.J.M. 
Horing, and X.L. Lei, cond-mat/0511148, in which the term relevant with 
$(\frac{W}{\chi})^2$ is stemming from the direct tunneling while the terms 
relevant with $(\frac{W}{\chi})$ results from the interference effect in 
Eq.~(10) for current and Eq.~(13) for shot noise.}

\bibitem{DongE}{We took no account of the contribution of the direct tunneling 
term in our previous paper, Ref.~\onlinecite{Dong3}. If this term is included, 
the linearly bias-voltage relevant term in the charge current formula, 
Eq.~(36), becomes $4V$ in stead of $3V$.}

\end{thebibliography}
\end{document}